\documentclass[a4paper,fleqn,usenatbib]{mnras}

\usepackage{newtxtext,newtxmath}
\usepackage{longtable}
\usepackage[T1]{fontenc}
\usepackage{ae,aecompl}

\usepackage{graphicx}	% Including figure files
\usepackage{amsmath}	% Advanced maths commands
\usepackage{amssymb}	% Extra maths symbols
\usepackage{ulem}
\usepackage{float}
\usepackage{placeins}

\title[VVV WIN 1733$-$3349]{VVV WIN 1733$-$3349:  a low extinction window  to probe the
  far side of the Milky  Way bulge \thanks{Based on observations taken
    within  the  ESO  Public  Surveys  VVV  and  VVVX,  Programme  IDs
    179.B-2002 and 198.B-2004, respectively.}}

\author[R.~K.~Saito et al.]{R.~K.~Saito$^{1}$\thanks{E-mail: saito@astro.ufsc.br},
D.~Minniti$^{2,3,4}$,
R.~A.~Benjamin$^{5}$,
M.~G.~Navarro$^{2,6,3}$,
\newauthor
J.~Alonso-Garc\'{i}a$^{7,3}$,
O.~A.~Gonzalez$^{8}$,
R.~Kammers$^{1}$,
F.~Surot$^{9}$
\\
\\
$^{1}$Departamento de  F\'{i}sica, Universidade  Federal de
  Santa Catarina, Trindade 88040-900, Florian\'opolis, SC, Brazil\\
$^{2}$Departamento  de  Fisica,  Facultad  de  Ciencias  Exactas,
  Universidad  Andres Bello,  Av.  Fernandez  Concha 700,  Las Condes,\\
  Santiago, Chile\\
$^{3}$Instituto Milenio de Astrof\'isica, Santiago, Chile\\
$^{4}$Vatican Observatory, V00120 Vatican City State, Italy\\
$^{5}$Department of Physics, University of Wisconsin-Whitewater, 800 W
  Main St, Whitewater, WI 53190 USA\\
$^{6}$ Dipartimento di  Fisica, Universit\`a degli Studi  di Roma ``La
Sapienza'', P.le Aldo Moro, 2, I00185 Rome, Italy\\
$^{7}$Centro  de Astronom\'ia  (CITEVA),  Universidad de  Antofagasta,
  Av. Angamos 601, Antofagasta, Chile\\
$^{8}$UK  Astronomy Technology  Centre,  Royal Observatory,  Blackford
  Hill, Edinburgh EH9 3HJ, UK\\
$^{9}$Instituto de Astrof\'{i}sica de  Canarias, E\--38205, La Laguna,
  Tenerife, Spain\\ 
}

\date{Accepted XXX. Received YYY; in original form ZZZ}

\pubyear{2015}

\begin{document}
\label{firstpage}
\pagerange{\pageref{firstpage}--\pageref{lastpage}}
\maketitle

% Abstract of the paper
\begin{abstract}

Windows of low  extinction in the Milky Way (MW)  have been used along
the  past decades  for the  study of  the Galactic  structure and  the
stellar population  across the inner  bulge and disk.  Here  we report
the analysis  of another low  extinction near-IR window  discovered by
the VISTA Variables in the V\'ia L\'actea Survey.  VVV WIN 1733$-$3349
is about  half a degree in  size and is conveniently  located right in
the MW  plane, at Galactic coordinates  $(l, b) = (-5.2,  -0.3)$.  The
mean extinction of VVV WIN  1733$-$3349 is $A_{Ks} = 0.61\pm0.08$ mag,
which is much smaller than the extinction in the surrounding area. The
excess in the  star counts is consistent with  the reduced extinction,
and complemented by studying the distribution of red clump (RC) stars.
Thanks to the strategic low-latitude  location of VVV WIN 1733$-$3349,
we  are able  to  interpret  their RC  density  fluctuations with  the
expected overdensities due  to the presence of the  spiral arms beyond
the  bulge.  In  addition, we  find a  clear excess  in the  number of
microlensing  events   within  the  window,  which   corroborates  our
interpretation that VVV  WIN 1733$-$3349 is revealing the  far side of
the MW bulge.

\end{abstract}


%%%%%%%%%%%%%%%%%%%%%%%%%%%%%%%%%%%%%%%%%%%%%%%%%%

%%%%%%%%%%%%%%%%% BODY OF PAPER %%%%%%%%%%%%%%%%%%

\section{Introduction}
\label{sec:intro}

Windows  of low  extinction in  the Milky  Way are  very important  in
Astronomy. A classical example, relevant  to the study of star forming
regions, is  the low optical  extinction window that allowed  the deep
exploration of the  innermost part of the Orion  Giant Molecular Cloud
revealing the secrets  of the Trapezium cluster and  imaging the first
protoplanetary  disks  with  the Hubble  Space  Telescope  \citep[HST;
][]{1994ApJ...436..194O,1996AJ....111..846O}.     Another    important
example is  Baade's window,  an optical  extinction window  of roughly
half a  degree in size  centred at  Galactic coordinates $(l,  b)= (1,
-4)$  deg, that  allowed the  study of  the bulge  stellar populations
\citep{1946PASP...58..249B,1963esag.book.....B,1985MmSAI..56...15B}
and it continues to  be used as a reference field  due to its detailed
characterisation (see \citet{2018ARA&A..56..223B} for a review).

Some other well known examples of low extinction windows used to study
Galactic   structure  and   stellar  populations   are:  Sgr   windows
\citep{1963esag.book.....B,1976MNRAS.174..169L},     Plaut's    window
\citep{1970A&A.....8..341P,1975A&A....41...71O,1982MNRAS.198..199G},
Sweeps     window     \citep{1998astro.ph..2307S,2008ApJ...684.1110C},
W0.2$-$2.1   and  W359.4$-$3.1   windows  \citep{2002A&A...381..219D},
Centaurus window \citep{2005A&A...433..931B}, etc.

Much  of what  we initially  learned about  the bulge  populations was
based on the  study of such low extinction windows  that allowed us to
see deep into the Galactic bulge \citep[e.g., ][]{1976MNRAS.174..169L,
  1978ApJ...226..777W,    1984AJ.....89..636B,    1988AJ.....96..884T,
  1996ApJ...460L..37S, 1999AJ....117.2296F} and  also to constrain the
star  count models  of  the  Milky Way  like  the Bensan\c{c}on  model
\citep[e.g., ][]{1986A&A...157...71R}.  Recent  studies of these bulge
windows carried out  with the HST have allowed to  explore the complex
star    formation   history    \citep[e.g.,   ][]{2011ApJ...735...37C,
  2016A&A...593A..82H, 2018MNRAS.477.3507B}.

The advent of large  IR surveys such as the Two  Micron All Sky Survey
\citep[2MASS;  ][]{2006AJ....131.1163S}, the  VISTA  Variables in  the
V\'ia  L\'actea  Survey  \citep[VVV; ][]{2010NewA...15..433M}  in  the
near-IR,  the  Spitzer  Galactic   Legacy  Infrared  Mid-Plane  Survey
Extraordinaire   \citep[GLIMPSE;  ][]{2003PASP..115..953B},   and  the
Wide-field       Infrared      Survey       Explorer      \citep[WISE;
][]{2010AJ....140.1868W}  in  the  mid-IR, enabled  to  make  detailed
reddening and extinction  maps, to obtain a more global  view of inner
Galactic structure  and stellar  populations, and  also to  search for
additional  windows   of  low  extinction  located   at  low  Galactic
latitudes.

Our  VVV Survey  in  particular  has recently  produced  a variety  of
extinction maps towards the Galactic bulge \citep{2011A&A...534A...3G,
  2018MNRAS.481L.130G,    2014AJ....148...24S,    2014A&A...571A..91M,
  2018A&A...616A..26M,    2018A&A...619A...4A,    2013A&A...552A.101S,
  2019MNRAS.488.2650S}. Using  these near-IR  extinction maps  we have
recently reported the discovery of VVV WIN 1713$-$3939 (a.k.a. Dante's
window),  located at  $(l, b)=  (347.4, -0.4)$  deg, that  allowed the
identification of  the spiral  arm structure  in the  far side  of the
Milky Way \citep{2018A&A...616A..26M}.

In  this  paper  we  present   the  characterisation  of  another  low
extinction near-IR window of roughly half a degree in size, located in
the Galactic plane at Galactic coordinates $(l, b)= (-5.2, -0.3)$ deg,
that exhibits not  only an overdensity in the star  counts, but also a
clear excess  of microlensing events  with respect to  its surrounding
regions.

\section{Observations}
\label{sec:obs}

\begin{figure}
\centering
\includegraphics[scale=0.8]{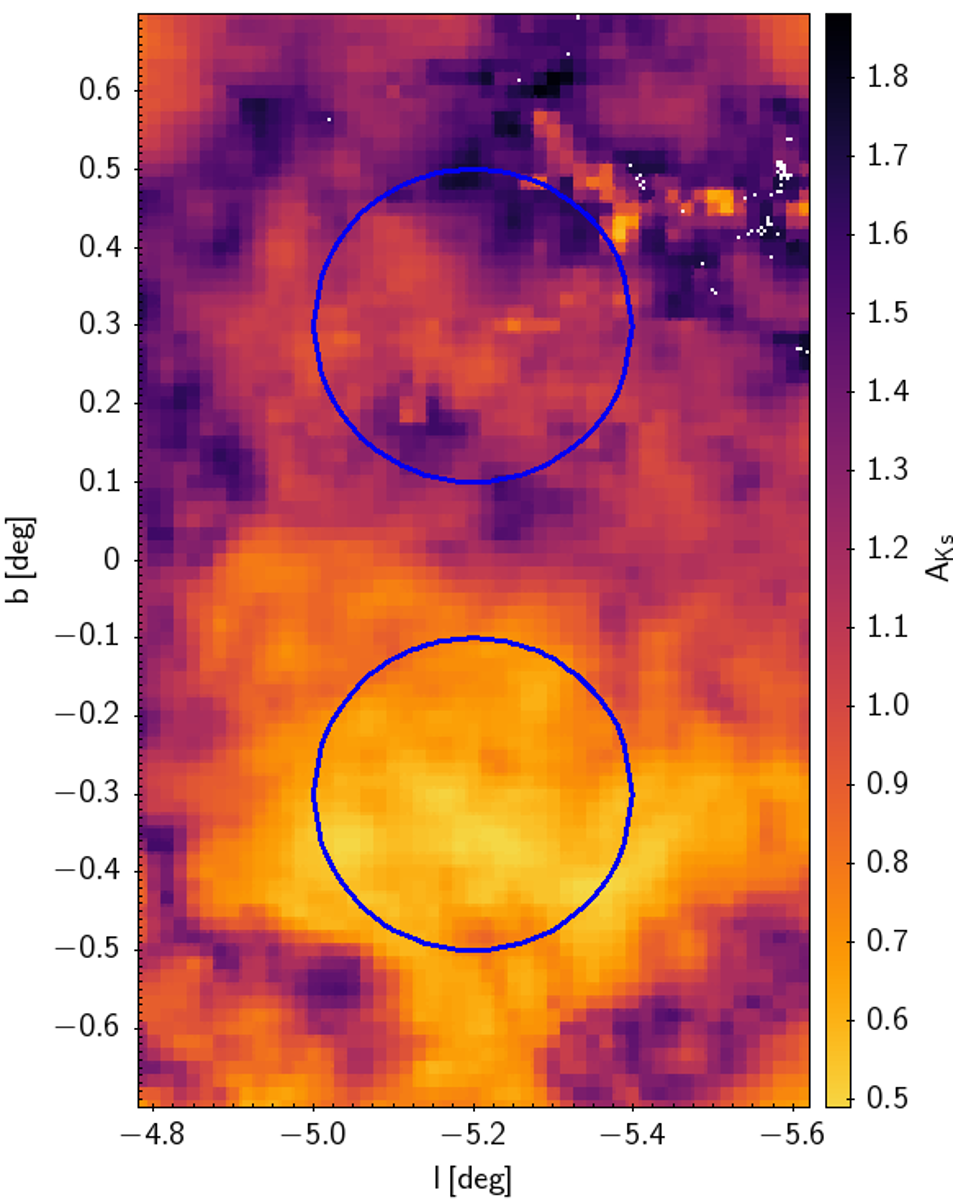}
\caption{A   modified   version  of   the   VVV   extinction  map   of
  \citet{2019A&A...629A...1S}   for   the   region  around   VVV   WIN
  1733$-$3349.  Two circles mark the regions defined as ``window'' (at
  negative latitudes) and ``control field'' (positive $b$).}
\label{fig:map}
\end{figure}

The ESO VISTA Variables in V\'{\i}a L\'actea (VVV) survey has recently
completed near-IR observations  of 562~sq.~deg.  area of  the MW bulge
and the  adjacent plane.  The  VVV strategy  consisted in two  sets of
quasi simultaneous $ZY$ and  $JHK_{\rm s}$ observations, plus $50-100$
observations  in $K_{\rm  s}$~band over  several years  ($2010-2015$),
providing a deep,  high-resolution dataset of the inner  Galaxy in the
near-IR \citep{2010NewA...15..433M,2012A&A...537A.107S}.

The standard  VVV photometry  is based on  aperture photometry  on the
stacked VVV  tile images, and  provided by the  Cambridge Astronomical
Survey  Unit  \citep[CASU;  for  details  see][]{2012A&A...537A.107S}.
More recently, point-spread function (PSF) photometry was performed on
the VVV images, and catalogues with  data in the different VVV filters
have been  released for the  inner regions of our  Galaxy \citep[e.g.,
][]{2018A&A...619A...4A, 2019A&A...629A...1S}. Due  to a high crowding
in the  inner bulge  -- where the  window is located  -- the  VVV data
presented here are based on PSF photometry.

\begin{figure*}
\centering
\includegraphics[scale=0.98]{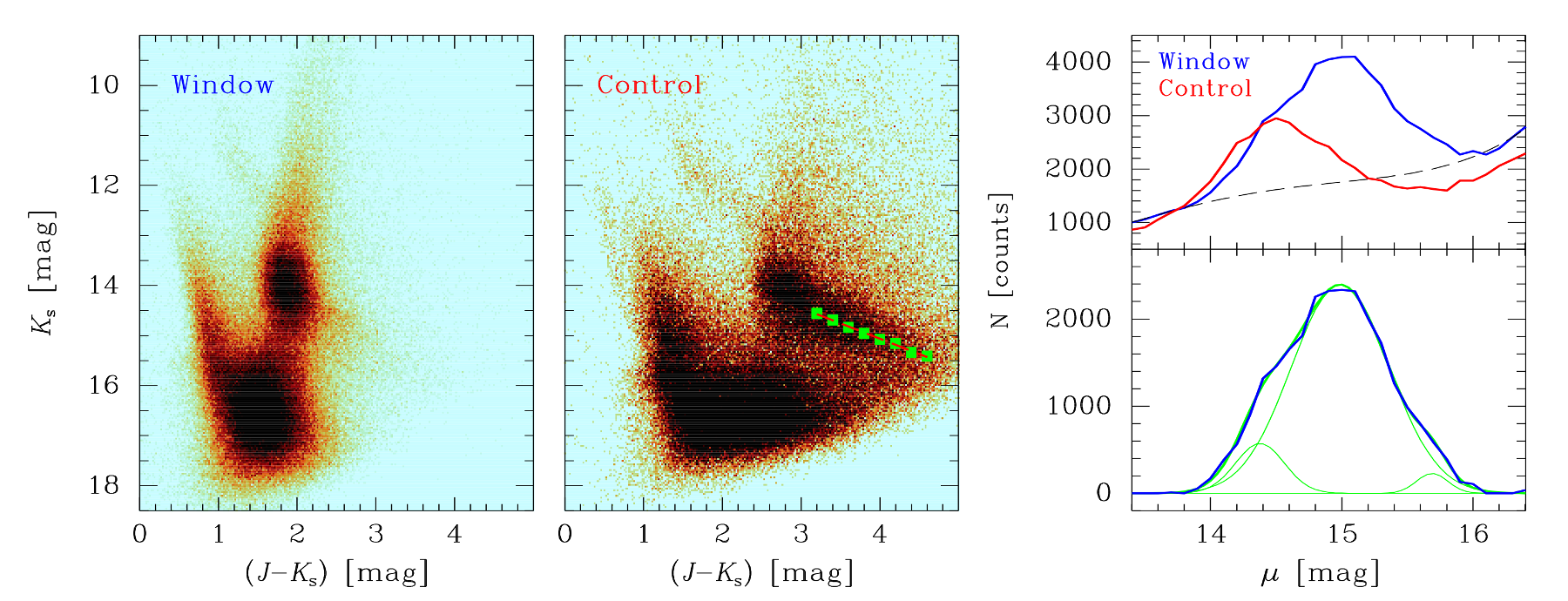}
\vspace{-0.1cm}
\caption{$K_{\rm  s}$ ~vs~  $(J-K_{\rm s})$  colour magnitude  diagram
  (CMD)   for    the   window    (left-panel)   and    control   field
  (central-panel). In the CMD for the control field a dashed line mark
  the  slope of  $A_{Ks} =  (0.612 \pm  0.018) \times  E(J-K_{\rm s})$
  measured for the reddening  vector. Top-right panel: distribution of
  selected  RC versus  distance moduli  of VVV  WIN 1733$-$3349  (blue
  line) and of  the control field (red line). The  polynomial fit used
  to subtract the  background LF of VVV WIN 1733$-$3349  is also shown
  with a dashed line.  Bottom-right  panel: multi Gaussian fit for the
  distribution of window  RC stars after subtracting  a polynomial fit
  to the  luminosity function. Information  about the three  peaks are
  listed in Table 1.}
\label{fig:cmds}
\end{figure*}

\section{Discussion}

\cite{2018MNRAS.481L.130G} made use of VVV red clump stars detected in
\citet{2018A&A...619A...4A}  catalogues  to create  a  high-resolution
reddening map of  the inner region of the Milky  Way bulge ($-10<l<10$
deg  and $-1.5<b<1.5$  deg).  They  discovered the  presence of  a low
extinction window  right in the  Galactic plane ($b\sim-0.3$  deg), at
just about  five degrees from  the Galactic centre  ($l\sim-5.2$ deg).
This window named VVV WIN 1733$-$3349 in \cite{2018MNRAS.481L.130G} is
also clearly  seen in the  maps of \citet{2019A&A...629A...1S}  and in
the distribution of Gaia DR2 sources \citep{2018A&A...616A...1G}.  The
region  is  about half  degree  in  size,  as  can be  appreciated  in
Fig.~\ref{fig:map},   which   shows   an  adapted   version   of   the
\citet{2019A&A...629A...1S} map for the region around the window.

In  our  analysis we  defined  as  the  window  a circular  region  of
24$\arcmin$  diameter   centered  at   $(l,  b)   =  (-5.2,-0.3)$~deg,
corresponding to RA, Dec (J2000)  = 17:33:51.92, $-$33:29:56.92 in the
equatorial system.   The position of WIN  1733$-$3349 (a.k.a.  Oscar's
window) is especially  important since for a  latitude of $b=-0.3$~deg
the vertical  projection along the line  of sight is small  ($z < 100$
pc) even at the far side of the MW disk ($d\lesssim 18$ kpc).

\subsection{Star counts and colour-magnitude diagrams}

Using  the  \citet{2009ApJ...696.1407N}  extinction law,  we  measured
$A_{Ks}  =  0.61 \pm  0.08$  mag  for  VVV  WIN 1733$-$3349  from  the
\citet{2019A&A...629A...1S}  extinction maps,  which  is much  smaller
than the  extinction in the  surrounding area.  For instance,  for the
symmetric area at  positive latitude (24$'$ diameter  area centered at
$l, b  = -5.2,+0.3$~deg.)  the extinction  is roughly twice as  in the
window,  with  $A_{Ks}=1.20  \pm  018$~mag.  This  symmetric  area  at
positive latitude  can be  used as  a control  field because  there is
negligible      disk      warping       at      these      coordinates
\citep[e.g.,][]{2006A&A...451..515M}.   Considering   similar  stellar
populations  in both  regions above  and below  the plane  (window and
control field,  respectively), the  differences observed  in comparing
them can be interpreted as caused by different extinction levels.

Fig.~\ref{fig:cmds} shows  the $K_{\rm  s} ~vs~ (J-K_{\rm  s})$ colour
magnitude diagram (CMD)  for VVV WIN 1733$-$3349  (left-panel) and the
control field (right-panel). The difference in the extinction level is
remarkable, with  the CMD  for the control  field stretched  along the
reddening vector.   While producing  the CMDs,  an examination  on the
stellar  density  shows  a  higher   density  of  sources  within  WIN
1733$-$3349  in  all   VVV  filters  when  compared   to  the  control
field. Indeed,  the stellar density in  the window is about  two times
larger than that  of the control field. The difference  in the stellar
density   is   even   larger   in    the   optical   Gaia   DR2   data
\citep{2018A&A...616A...1G}, with  about four times more  stars within
the window compared with the control field. Similar ratio is also seen
in the optical Pan-STARRS1 \citep{2016arXiv161205560C} and DECaPS data
\citep{2018ApJS..234...39S}.

\begin{figure}
\centering
\includegraphics[scale=0.83]{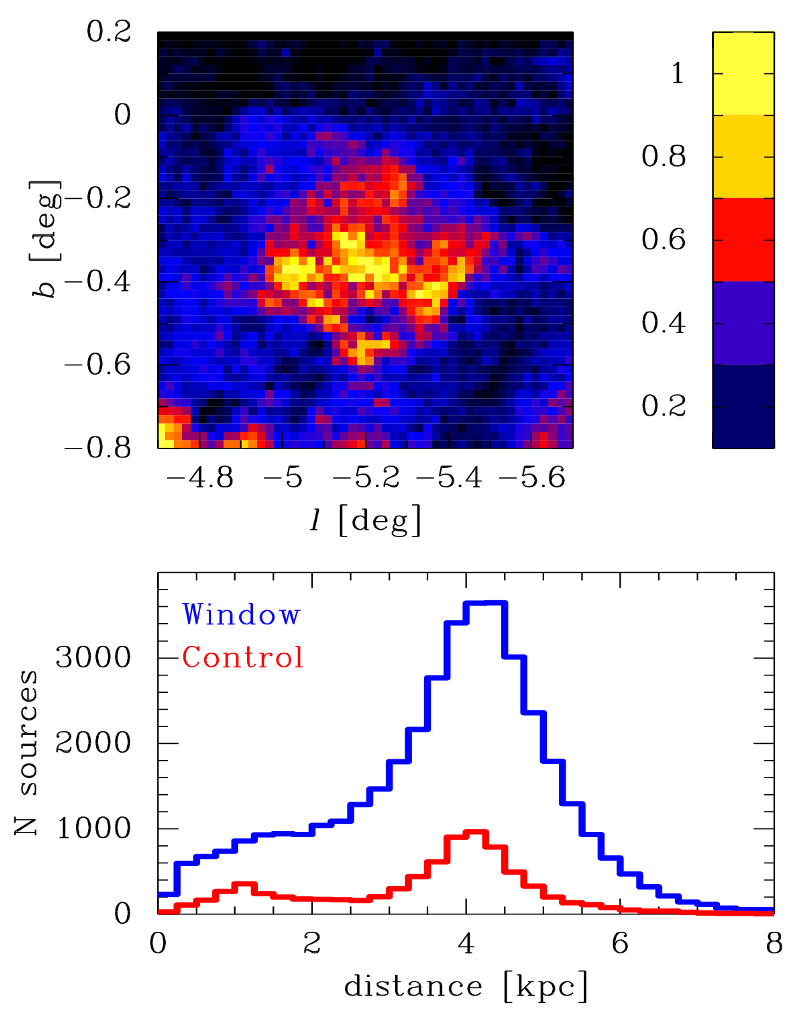}
\vspace{-0.1cm}
\caption{Top panel: density map of  Gaia DR2 ($G$ band) sources around
  the window normalized by the  maximum density value.  A vertical bar
  shows the  colour code  in the map.   Bottom panel:  distribution in
  distance    for    the    Gaia     DR2    sources    according    to
  \citet{2018AJ....156...58B}.}
\label{fig:count}
\end{figure}

Fig.~\ref{fig:count} presents in  the mid panel a density  map for the
optical Gaia DR2 data.  The  spatial distribution shows an overdensity
in agreement with  the location and size of the  of WIN 1733$-$3349 as
seen in the extinction map (see Fig.~\ref{fig:map}).  The distribution
in distance for  the Gaia DR2 sources within  WIN 1733$-$3349 compared
with    the   control    field   \citep[from    ][bottom   panel    of
  Fig.~\ref{fig:count}]{2018AJ....156...58B}   shows  that   the  Gaia
photometry  is not  deep enough,  reaching barely  to the  bulge stars
($\sim 7$~kpc).

\begin{figure*}
\centering
\includegraphics[scale=0.76]{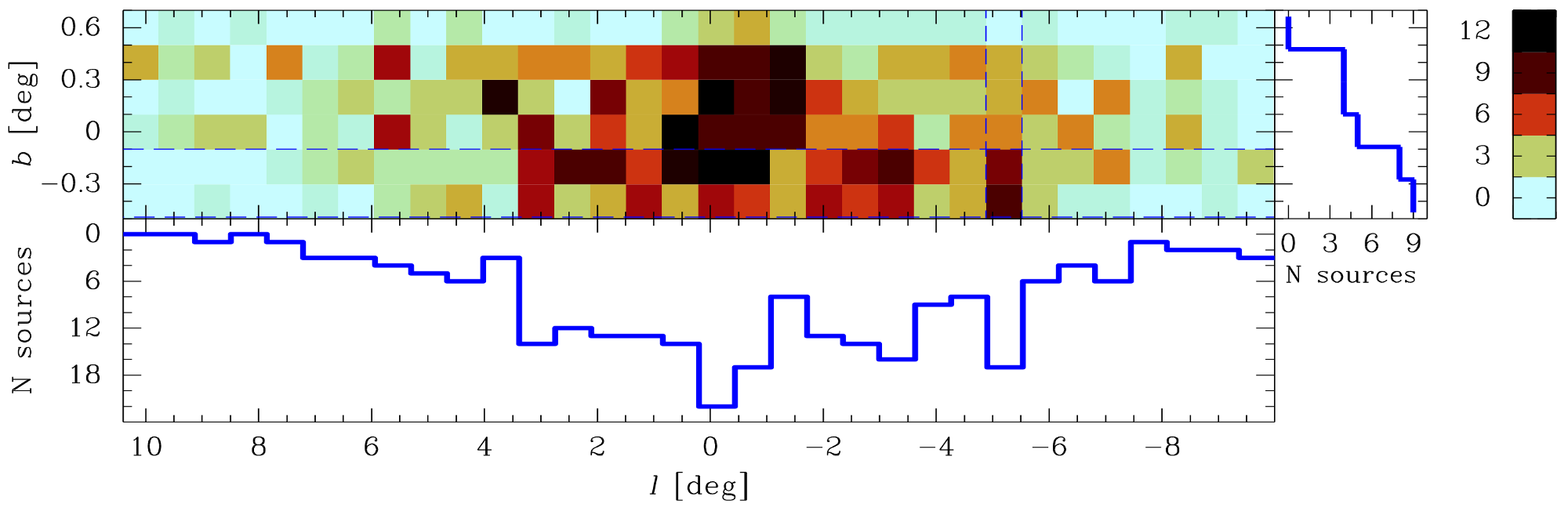}
\vspace{-0.1cm}
\caption{Density plot with the  distribution of microlensing events in
  the  inner  bulge.  An  overdensity  of  events  is present  at  the
  position  of  VVV  WIN  1733$-$3349   ($l,  b  =  -5.2,  -0.3$~deg).
  Histograms  for   sources  within   the  WIN   1733$-$3349  limiting
  coordinates (dashed lines) are also  shown for both axes. A vertical
  bar   shows   the   colour   code  in   the   map.    Adapted   from
  \citet{2018ApJ...865L...5N,2019arXiv190704339N}.}
\label{fig:lenses}
\end{figure*}

We can explore zones with more transparency towards the inner MW bulge
in order to map structures towards -- or even behind -- the bulge. For
this we made use of red  clump stars as distance indicators within VVV
WIN 1733$-$3349.   We selected red clump  stars using a simple  cut in
colour  in the  CMDs and  then  calculated the  distance assuming  the
recently calibrated  absolute magnitude of  RC stars provided  by Gaia
DR2 as  $M_{Ks}= -1.605  \pm 0.009$  and $(J-K_{\rm  s})_0 =  0.66 \pm
0.02$ mag  \citep{2018A&A...609A.116R} and the slope  of the reddening
vector measured  directly from the  control field CMD which  yields to
$A_{Ks} = (0.612 \pm 0.018) \times E(J-K_{\rm s})$.  With those values
the  distance modulus  for  the RC  stars  is  given by  $\mu  = -5  +
5\,log\,d  (pc)=K_{\rm s}-0.612\times(J-  K_{\rm s})+2.009$~mag.   The
slope of the  reddening vector is flatter than $A_{Ks}  = 0.725 \times
E(J-K_{\rm  s})$ from  \citet{1998ApJ...500..525S},  but steeper  than
$A_{Ks}      =     0.528      \times     E(J-K_{\rm      s})$     from
\citet{2009ApJ...696.1407N}. These  values are  much steeper  than the
slope in  \citet{2017ApJ...849L..13A} and \citet{2018A&A...616A..26M},
where  $A_{Ks} =  0.428 \times  E(J-K_{\rm  s})$ and  $A_{Ks} =  0.484
\times  E(J-K_{\rm   s})$,  respectively.  The  use   of  a  different
extinction law causes the distance scale to shrink/expand.

The luminosity function (LF) for the WIN 1733$-$3349 RC stars compared
with the  control field RC  stars is shown  in the top-right  panel of
Fig.~\ref{fig:cmds}.  The distributions are quite different, with many
more RC stars in the window LF, which peaks much farther when compared
with  the control  field and  presents a  composite distribution.   In
order to subtract the background and  enhance possible peaks of the RC
along  the line  of sight,  In order  to subtract  the background  and
enhance possible peaks of the RC along the line of sight, we applied a
polynomial $+$ multi Gaussian fit,  as shown in the bottom-right panel
of Fig.~\ref{fig:cmds}.  The resulting peaks  due to the  different RC
are   fitted  with   a   Levenberg-Marquardt  optimization   algorithm
\citep[e.g.,  ][]{2004...172_2K}   in  order  to  obtain   their  mean
distances  and respective  standard  deviations, which  are listed  in
Table 1.

At the magnitude range of the fainter RC stars the completeness of the
VVV   PSF  data   on  Galactic   plane  is   over  80\%   \citep[e.g.,
][]{2012A&A...537A.107S,2019A&A...629A...1S}.       Following      the
discussion  presented in  \cite{2018MNRAS.481L.130G},  we assume  that
secondary peaks  in the  RC distributions  are not  caused by  the red
giant                            branch                           bump
\citep[e.g.,][]{2011ApJ...730..118N,2013ApJ...766...77N}.  In contrast
with    the    distances   from    Gaia    DR2    as   estimated    by
\citet{2018AJ....156...58B}, the  distances to the RC  giants measured
from the VVV near-IR photometry clearly penetrates through the Galaxy,
out  to $\sim  14$  kpc.   Using a  flatter  reddening  vector in  our
calculations scales the RC distribution  even to larger distances. For
instance,   with  $A_{Ks}   =  0.428   \times  E(J-K_{\rm   s})$  from
\citet{2017ApJ...849L..13A} the  distances to  the RC are  $\sim 15\%$
larger. This difference in slopes  with previous VVV measurement could
be due to the small number of RC stars at a given color in our control
field, which, along with the relative closeness of the RC of the bulge
and of the  background spiral arm at these longitudes  (Minniti et al,
in prep.), complicates its accurate measurement.

\subsection{Microlensing}

\citet{2018ApJ...865L...5N,    2019arXiv190704339N}    searched    for
microlensing events  in bulge  fields along  the Galactic  plane. They
discovered several hundreds  of events in a large  area covering about
21  sq.   deg.   (within  $-10.00<l<+10.44$  deg,  and  $0.46<b<+0.65$
deg). The analysis  of the spatial distribution  of these microlensing
events revealed an unexpected excess  in the surface density of events
at  $(l, b)=(-5.2,  -0.3)$ deg,  in an  area that  corresponds to  the
location  of the  present  window  (see Fig.~\ref{fig:lenses}).   This
excess in the number density of events above the neighboring area is a
3 sigma result.

The traditional  microlensing configurations  (bulge-bulge, bulge-disk
and disk-disk events)  should produce an excess of events  in the very
center of the MW due  to the higher stellar density \citep{navarro17},
but not in this area. The  most straightforward explanation is that we
are seeing all the way  through the bulge, including additional events
that have lenses in  the bulge and sources in the  far disk. The first
event with this new far  disk-bulge event configuration was discovered
by the  United Kingdom Infrared  Telescope Survey (UKIRT) at  $(l,b) =
(-0.12,-0.33)$  \citep{yossi17,yossi18}.   Another candidate  for  far
disk event at  very low latitude was reported  by \cite{bennett18b} at
$(l,  b)=(0.90,-1.97)$.   A  more  comprehensive  study  seems  to  be
confirmed by the larger estimated distances  of the RC sources in this
region (Navarro  et al.  2020,  in preparation).  We take  this excess
number  of microlensing  events in  this area  as additional  evidence
supporting the conclusion that VVV WIN 1733$-$3349 is a is a window of
low near-IR extinction piercing deep through the entire bulge region.

\begin{table}
\centering
\begin{tabular}{ccl}
\hline
\hline
RC peaks  & RC peaks    & Interpretation  \\
$\mu$ ($K_{\rm s}$) [mag] & Dist. [kpc]  &               \\
\hline
$14.376 \pm 0.373$ ($\sim$13.51) & \,\,7.50 & Scutum-Centaurus arm\\ 
$14.988 \pm 0.742$ ($\sim$14.13) & \,\,9.94 & Galactic bar \\ 
$15.689 \pm 0.236$ ($\sim$14.85) & 13.73 & Sagittarius arm\\
\hline
\hline
\end{tabular}
\caption{Distances  for  the RC  peaks  from  the multi  Gaussian  $+$
  polynomial fit to the window RC luminosity function.}
\vspace{-0.1cm}
\end{table}

\section{Conclusions}

We  present the  further  characterisation of  another  window of  low
extinction in the Galactic bulge, VVV WIN 1733$-$3349 (a.k.a.  Oscar's
window).  This window is strategically located in the plane, at $(l,b)
= (-5.2, -0.3)$ deg, and has a diameter of about 24$'$.  The window is
clearly seen in the  VVV extinction maps for the region  as well as in
the distribution of Gaia DR2 sources.

We  use  the   deep  VVV  PSF  photometry  to   measure  the  distance
distribution of RC giants along the  line of sight.  The RC giants can
be seen out to $\sim 14$ kpc, in the far disk well beyond the bulge. A
multi Gaussian fit  shows the presence of three  peaks, interpreted as
the Scutum-Centaurus arm in the  foreground disk, the Galactic bar and
the Sagittarius arm (that is  seen wrapping around behind the Galactic
centre),  at  the  distance  of  $d=7.06$,  $9.37$  and  $12.92$  kpc,
respectively.  The  distance of the  Galactic bar and  the Sagittarius
arm   coincide    with   distances   for   RC    stars   measured   by
\citet{2018MNRAS.481L.130G} for the region within $|b|<1.5$~deg.

We also find a clear excess of microlensing events in this window from
the sample  of \citet{2018ApJ...865L...5N,  2019arXiv190704339N}, that
is probably due to additional events that have lenses in the bulge and
sources in the far disk. That  is in agreement with our interpretation
of the RC distribution, revealing the far side of the Milky Way bulge.

\section*{Acknowledgements}

We gratefully acknowledge  the use of data from the  ESO Public Survey
program IDs 179.B-2002 and 198.B-2004  taken with the VISTA telescope,
and data products from the  Cambridge Astronomical Survey Unit (CASU).
This  publication  makes use  of  data  products from  the  Wide-field
Infrared Survey Explorer,  which is a joint project  of the University
of    California,    Los    Angeles,   and    the    Jet    Propulsion
Laboratory/California Institute of Technology,  funded by the National
Aeronautics  and Space  Administration.  R.K.S.   acknowledges support
from  CNPq/Brazil through  projects  308968/2016-6 and  421687/2016-9.
Support for  the authors is provided  by the BASAL CONICYT  Center for
Astrophysics   and  Associated   Technologies  (CATA)   through  grant
AFB-170002,  and  the  Ministry  for  the  Economy,  Development,  and
Tourism,  Programa  Iniciativa   Cient\'ifica  Milenio  through  grant
IC120009,  awarded   to  the  Millennium  Institute   of  Astrophysics
(MAS).  D.M.   acknowledges  support  from  FONDECYT  through  project
Regular  \#1170121.

% Don't change these lines
\bsp	% typesetting comment
\label{lastpage}
\end{document}